\documentclass{jfm}

\usepackage{graphicx}
\usepackage{natbib}
\usepackage{sidecap}

\usepackage{amsfonts}          
\usepackage{amssymb}           
\usepackage{amsmath}   
        
\usepackage{color}

\ifCUPmtlplainloaded \else
  \checkfont{eurm10}
  \iffontfound
    \IfFileExists{upmath.sty} 
      {\typeout{^^JFound AMS Euler Roman fonts on the system,
                   using the 'upmath' package.^^J}%
       \usepackage{upmath}}
      {\typeout{^^JFound AMS Euler Roman fonts on the system, but you
                   dont seem to have the}%
       \typeout{'upmath' package installed. JFM.cls can take advantage
                 of these fonts,^^Jif you use 'upmath' package.^^J}%
      }
  \else
  \fi
\fi

\ifCUPmtlplainloaded \else
  \checkfont{msam10}
  \iffontfound
    \IfFileExists{amssymb.sty}
      {\typeout{^^JFound AMS Symbol fonts on the system, using the
                'amssymb' package.^^J}%
       \usepackage{amssymb}%

      }{}
  \fi
\fi

\ifCUPmtlplainloaded \else
  \IfFileExists{amsbsy.sty}
    {\typeout{^^JFound the 'amsbsy' package on the system, using it.^^J}%
     \usepackage{amsbsy}}
    {}
\fi

\providecommand\calO{{\cal O}}

\newsavebox{\astrutbox}
\sbox{\astrutbox}{\rule[-5pt]{0pt}{20pt}}

\newcommand*{\Ray}{{\rm Ra}}
\newcommand{\Rob}{{\rm q}}
\renewcommand*{\Pr}{{\rm Pr}}
\newcommand*{\Pm}{{\rm Pm}}
\newcommand*{\Ek}{{\rm E}}
\newcommand*{\Rm}{{\rm Rm}}

\newcommand*{\Elsclass}{{\mbox{$\Lambda$}}}
\newcommand*{\Elsnum}{{\mbox{$\Lambda '$}}}

\def\bfnabla{\mbox{\boldmath $\nabla$}}

\def\bfu{\mbox{\bf u}}
\def\bfB{\mbox{\bf B}}
\def\bfe{\mbox{\bf e}}
\def\bfr{\mbox{\bf r}}

\def\dd{{\rm d}}

\title[Strong-Field Spherical Dynamos]{Strong-Field Spherical Dynamos}
\author[E. Dormy]{
E\ls M\ls M\ls A\ls N\ls U\ls E\ls L\ns D\ls O\ls R\ls M\ls Y,}

\affiliation{MAG (CNRS/IPGP/ENS), Ecole Normale Sup\'erieure, 24 rue
  Lhomond, 75005 Paris}

\date{\today}

\begin{document}

\maketitle
             
\begin{abstract}
Numerical models of the geodynamo are usually classified in two categories: those
denominated dipolar modes, observed when the inertial term is small enough,
and multipolar fluctuating 
dynamos, for stronger forcing.
We show that a third dynamo branch corresponding to a dominant force balance
between the Coriolis force and the Lorentz force can be produced
numerically. This force balance is
usually referred to as the strong-field limit. 
This solution co-exists with the often described viscous branch.
Direct numerical simulations
exhibit a transition from a weak-field dynamo branch, in which viscous
effects set the dominant length scale, and the strong-field branch
in which viscous and inertial effects are largely negligible.
These results indicate that a distinguished limit needs to be sought to produce
numerical models relevant to the geodynamo and that the usual approach of minimizing
the magnetic Prandtl number (ratio of the fluid kinematic viscosity to its
magnetic diffusivity) at a given Ekman number is misleading.
\end{abstract}

\begin{keywords}
geodynamo, geophysical and geological flows, magnetohydrodynamics.
\end{keywords}

\section{Introduction}

The origin of the Earth's magnetic field is a challenging problem.
It is now widely accepted that this magnetic field is generated by
an internal self-excited dynamo action in the conducting liquid core of the
Earth -- \cite[see][for an introduction]{INTRO_DYN,INTRO_DYN2}. 
Thermal energy is converted to kinetic energy via convective motions, 
which in turn are able to amplify electrical currents, and part of the
kinetic energy can thus be converted to 
magnetic energy. The amplification of electrical currents in the conducting fluid
is then saturated by the back-reaction of the Lorentz force on the flow.   
The nature of the transition from a purely hydrodynamic 
(non-magnetic)
solution to the dynamo solution as well as the saturation mechanisms 
remain largely open questions.

The geodynamo problem involves
the resolution of a set of fully nonlinear coupled equations describing
magnetohydrodynamics in a rotating reference frame. In the rapid rotation
limit, the system of governing equations 
becomes stiff and cannot be handled numerically as such. For this reason
all numerical simulations are, despite the use of state-of-the-art
computational resources, performed in a parameter regime far off
the relevant values.
This stiffness of the equations is directly related to extreme values taken by
ratios of typical time scales or typical length scales in the problem. In numerical
simulations, however, the controlling parameters assume much more moderate
values, and the corresponding time scales or length scales are necessarily
harder to distinguish. 

Most numerical models rely on the Boussinesq approximation (incompressible
fluid, except inasmuch as the buoyancy force is concerned) and rely on an
imposed temperature gradient across the Earth's core to drive thermal
convection.
Such numerical models, produced to date, appear to fall in two categories. For
moderate values of the control parameter, the Rayleigh number,
the produced magnetic field is largely dipolar axial, similar in that
respect to the Earth's magnetic field. It can exhibit time variations, but does not
reverse polarity \citep{Kutzner,Christensen1}.
At larger values of the Rayleigh number, a secondary bifurcation occurs
leading to a ``multipolar'' and fluctuating dynamo phase \citep{Kutzner}.

The nature of the dynamo onset (i.e. the bifurcation from the purely
hydrodynamic state to the first dynamo mode) has been studied in detail in
\cite{MD09}. We reported supercritical, subcritical and isola bifurcation diagrams
depending on the values of the parameters.
{A mechanism for the subcritical bifurcations, in terms of
  helicity enhancement, has been proposed by \cite{SreenivasanJ11}
 \citep[see also][]{Dormy2011}.}
The transition at larger forcing between the dipolar and
multipolar phases has been identified as being controlled 
by the relative strength of the curl of inertial forces to that of either
the viscous or the Coriolis term \citep[see][]{OrubaD14b}.

These two branches have also been reported in the presence of a uniformly
heated fluid as mean dipole (MD) and fluctuating dipole (FD) \citep{Simitev2009}.
The hysteretic nature of this transition and the existence of a domain of
bistability has been stressed by many authors
\citep{GoudardD08,Simitev2009}. \cite{Schrinner2012} show that 
in the presence of stress-free boundary conditions, the same transition occurs, and that the strong
hysteresis is associated with the particular nature of geostrophic flows. A similar behaviour is
to be expected in the presence of rigid boundary conditions, when viscous
effects are small enough.

The present paper focuses on a different mode, characterised by a regime
in which both inertia and viscosity are negligible, and the Lorentz force
relaxes the constraints imposed by rapid rotation.


\section{Governing equations}

Thermal convection and magnetic field generation in the Earth's core are
modelled in the present study using the most classical set of equations. The
rotating incompressible MHD equations are 
coupled to the heat equation under the Boussinesq approximation.
Convection is driven by an imposed
temperature difference across a spherical shell (of inner radius $r_i$
and outer radius $r_o$).
Magnetic field generation by dynamo action requires a flow
with an appropriate geometry and sufficient amplitude, which can be achieved if
the control parameter 
(measuring the efficiency of the thermal driving) 
is increased away from the onset of convection.
The parameter space for such dynamos has been extensively
studied by Christensen and collaborators 
\citep[e.g.][]{Christensen1} providing a detailed description of the ``phase diagram''
for dynamo action in this set-up (i.e. the region in the parameter space for which
different dynamo solutions are produced).
The governing equations are solved in a spherical shell 
($r_i/r_o=0.35$) and in a rotating reference frame. The 
reference frame is such that the velocity vanishes
on both spheres (no-slip boundaries), a temperature difference is
maintained for all time across the shell, and both the inner and the outer
domains are assumed electrically insulating.
The equations governing the
velocity, $\bfu$, magnetic field $\bfB$
and temperature field $T$ are then in non-dimensional form 
\begin{eqnarray}
\frac{\rm E}{\rm Pm} \, 
\left[\partial _t \bfu +  (\bfu \cdot \bfnabla) \bfu \right]
&=& 
- \bfnabla \pi 
+ {\rm E} \, \Delta \bfu
- 2 \bfe_z \times \bfu 
+ {\rm Ra\, q} \, T \, \bfr 
+ \left(\bfnabla \times \bfB \right) \times \bfB\, , \,\,\,\,\,\,
\label{NS}
\end{eqnarray}   
\begin{equation}
\partial _t \bfB = \bfnabla \times \left(\bfu \times \bfB
- \bfnabla \times \bfB \right) \, ,
\qquad
\partial_t T + (\bfu \cdot \bfnabla) T
= {\rm q}\, \Delta T\, ,
\end{equation} 
\begin{equation}
\mbox{with}\quad \bfnabla \cdot \bfu =
\bfnabla \cdot \bfB = 0 \, .
\end{equation} 
In the above equations $L=r_o-r_i$ has been used as length scale, $\tau_\eta=L^2/\eta$ as time
scale, $(\Omega \mu_0 \rho_0 \eta)^{1/2}$ as 
magnetic field scale. This non-dimensional form is well suited for
strong-field dynamos \citep[e.g.][]{Fearn1998}.  
The following non-dimensional numbers have been introduced: the Ekman number ${\rm E}={\nu}/{\Omega L^2}\, ,$
the magnetic Prandtl number ${\rm Pm}={\nu}/{\eta}\, ,$
the Roberts number ${\rm q}=\kappa/\eta \, ,$
and
the Rayleigh number ${\rm Ra}={\alpha g \Delta T L}/{\kappa \Omega}\, , $
where $g=g_o/r_o$ with $g_o$ the gravity at $r=r_o$ (note that the
Rayleigh number is here modified from its most standard definition to
account for the stabilizing effect of rotation).
It is also useful to define
the Prandtl number ${\rm Pr}={\nu}/{\kappa} \equiv {\rm Pm}/{\rm q}\, . $
In the present work, $\Ek$ is set to $3\, 10^{-4}$ and ${\rm Pr}$ to unity,
it follows that ${\rm q} = \Pm$ in the sequel. 

These equations are numerically integrated using the {\it Parody}
code, originally developed by the author and improved with several
collaborators~\citep[see][]{Parod1,Schrinner2012}.  {The
  numerical resolution in the simulations reported here is $132$ grid
  points in radius, with spherical harmonic decomposition of degrees up
  to $\ell_{\rm max}=256$ and modes up to $m_{\rm max}=64$. The models
  were integrated for up to $10$ magnetic diffusion times.}  In order
to ensure the validity of the new solutions presented here, these
simulations were also kindly reproduced by V.~Morin using the {\it
  Magic} code, developed by G.~Glatzmaier and modified by
U.~Christensen and J.~Wicht.  Both codes have been validated through
an international benchmark \citep[see][]{bench}.

\section{Weak- and strong-field dynamos}

\begin{figure*}
\centerline{\includegraphics[width=0.47\textwidth]{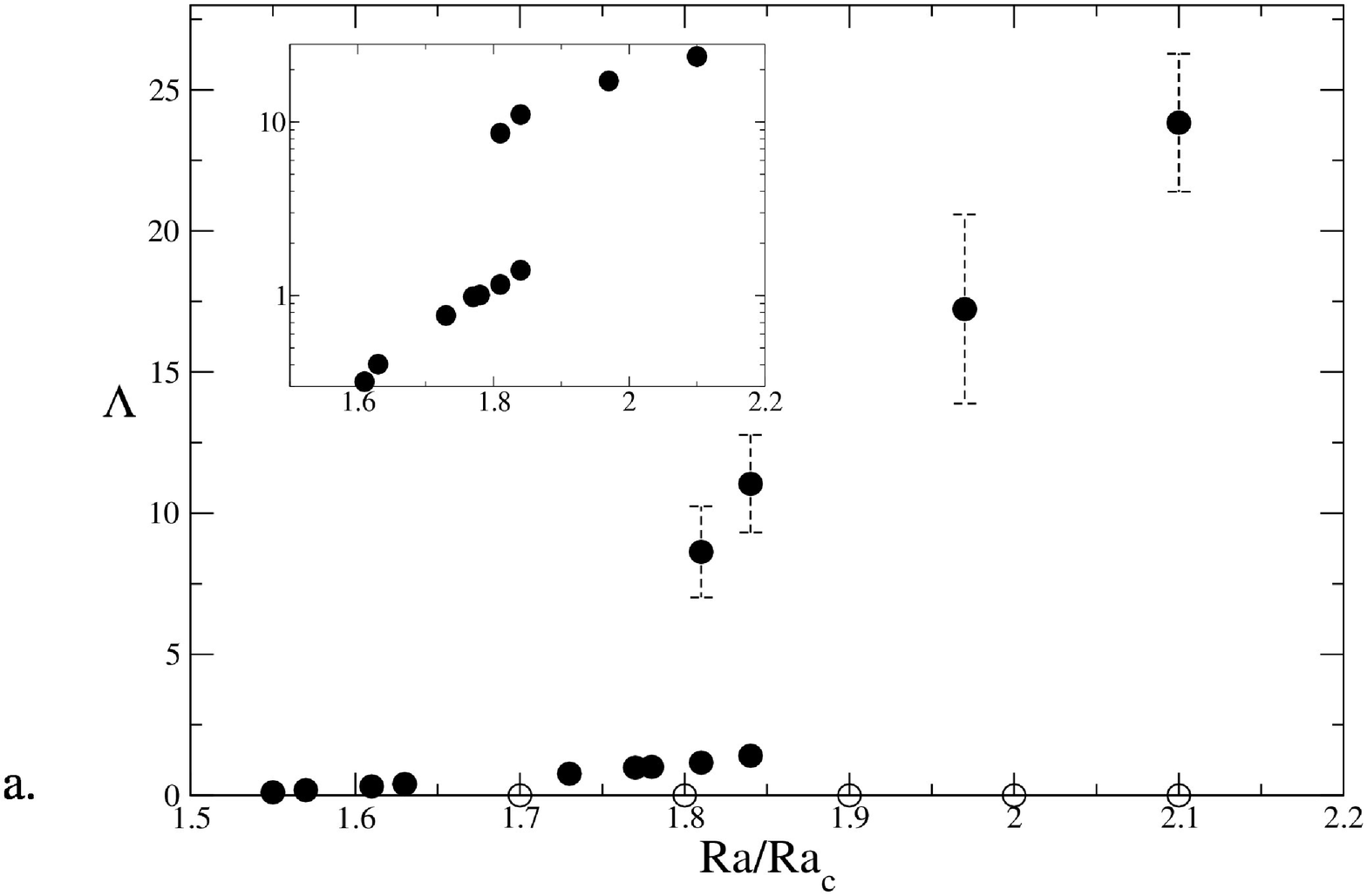}
\hskip6mm\includegraphics[width=0.47\textwidth]{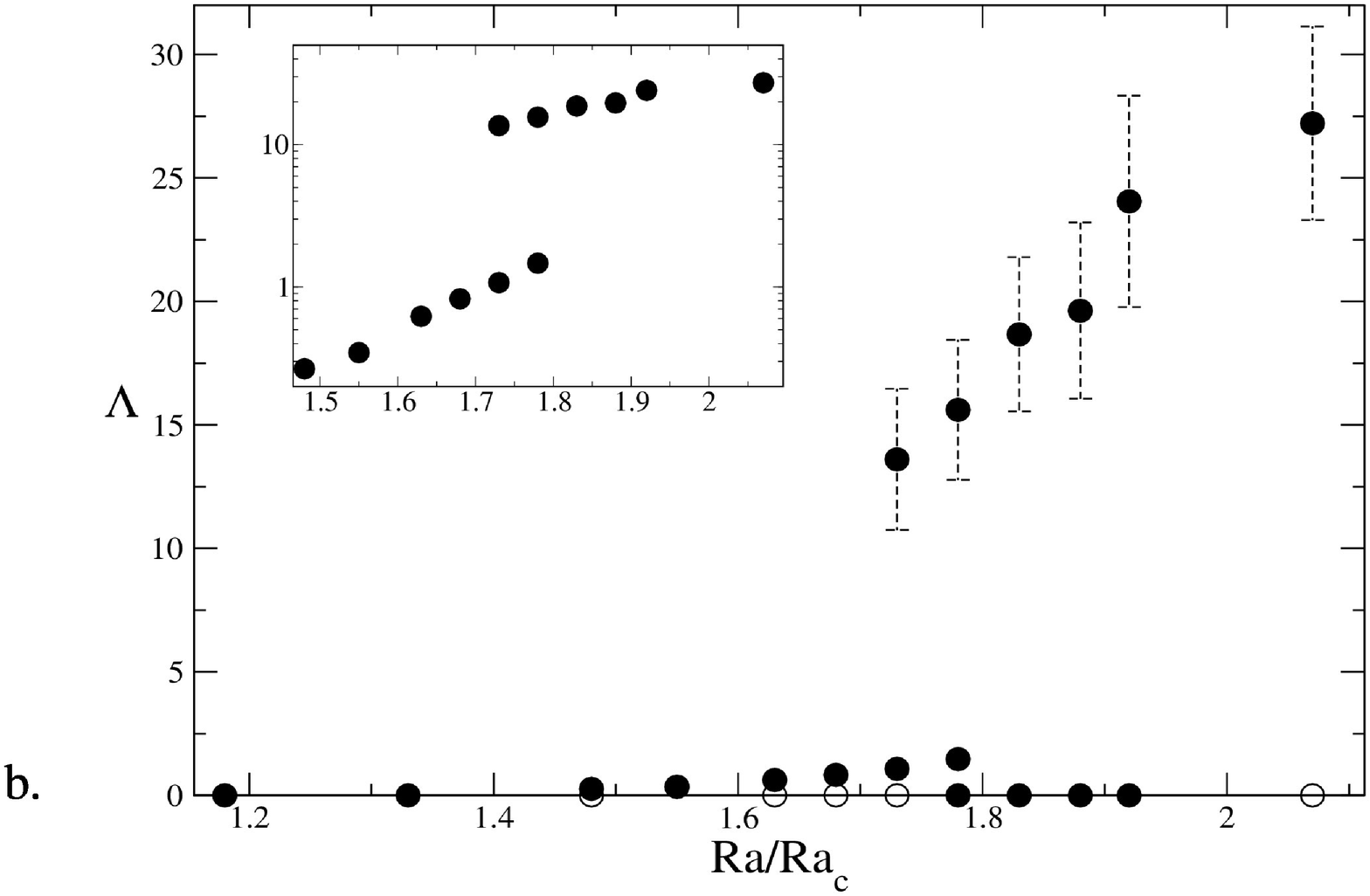}}
\caption{Weak- and strong-field branches for $\Ek=3 \cdot 10^{-4}$ and
{\bf a:} $\Pm=14$, and {\bf b:} $\Pm=18$. Symbol indicate the time averaged
Elsasser number ($\bullet$ stable, $\circ$
unstable), time variability of the dynamo mode 
is reported using the standard deviation (indicated with error bars).
The Insets present the same graphs in lin-log scale, so that the weak-branch
is more clearly visible.}
\label{fig_bif3a}
\end{figure*}

\begin{figure*}
\centerline{{\small a.}\includegraphics[width=0.4\textwidth]{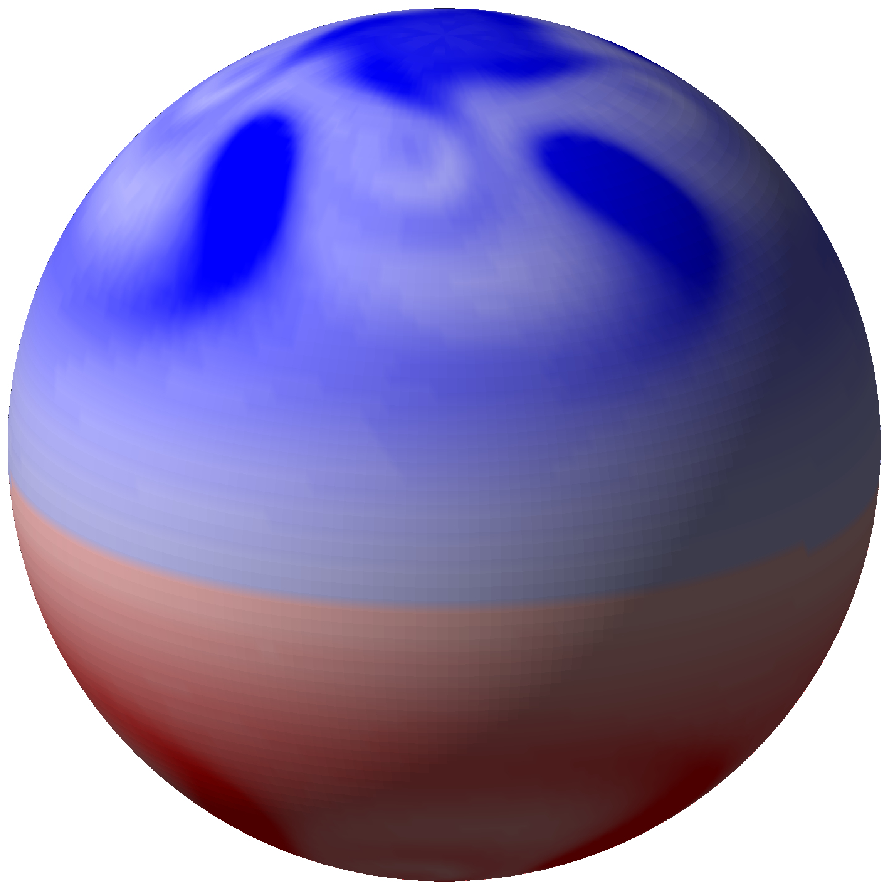}
\hskip6mm{\small b.}\includegraphics[width=0.4\textwidth]{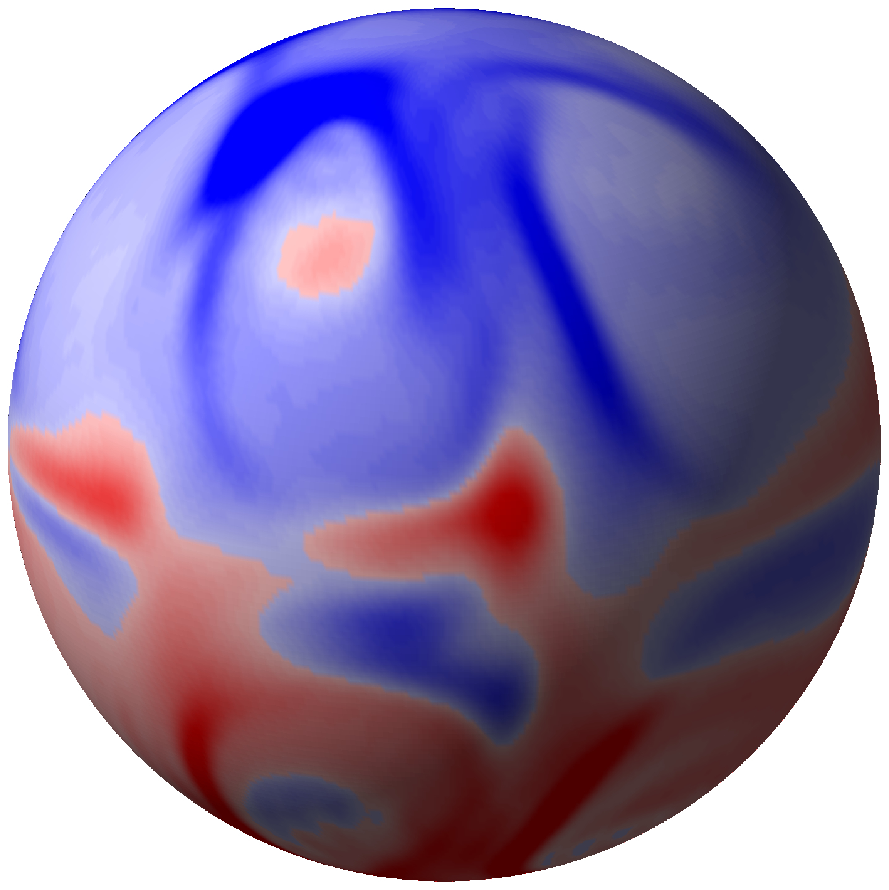}}
\caption{{Radial component of the magnetic field produced at the surface of the model for $\Pm=18$, and $\Ray/\Ray_c=1.73$ on the weak-field (a.) and strong-field (b.) branches. The field is characterised by a strong axial dipolar component on both branches.}}
\label{fig_Br}
\end{figure*}

Following the same approach as \cite{MD09}, we study the bifurcation from
the purely hydrodynamic solution to the dynamo state using the
Rayleigh number as control parameter.
At fixed E and Pm,
the Rayleigh number needs to exceed a given value for a dynamo solution 
(non-vanishing field) to exist. 
We report here direct numerical simulations performed at large values of the magnetic
Prandtl number Pm. One may object that such parameter regime is irrelevant to dynamo
action in liquid metals (characterised by a small magnetic Prandtl number).
We will however argue that considering large values of Pm can compensate for
the excessive role of inertial terms in numerical dynamo models, 
and is a necessary consequence of the large values assumed by the Ekman number.

Figure~\ref{fig_bif3a} presents the bifurcation diagrams obtained for ${\rm Pm}={\rm
  q} = 14$ and ${\rm Pm}={\rm q}=18$.
The magnetic field strength, as measured by the classical Elsasser number
\(
\Elsclass = {B ^2}/(2 \, \Omega \, \rho \, \mu \, \eta)
\)
is represented versus the Rayleigh number, normalised by its 
value at the onset of thermal convection $\Ray_c$ (here $\Ray_c=60.8$).
Each point on this figure corresponds to a time averaged fully
three-dimensional simulation. 
The time variability of the dynamo mode 
is reported using the standard deviation.

Figure~\ref{fig_bif3a} is characterised by a supercritical bifurcation (as reported in
\cite{MD09} for their ``large'' values of the magnetic Prandtl
number, i.e. $\Pm=6$).
However this first dynamo branch rapidly destabilises to a second branch of
much stronger amplitude. This strong-field branch can be maintained for
decreasing values of the Rayleigh number. 
The magnetic field is dominated by the axial dipole on both
branches {(see Figure~~\ref{fig_Br})}. 
The strong-field branch on Fig~\ref{fig_bif3a}a 
is hysteretic to the onset of dynamo itself: once on this
branch, the control parameter can be decreased below the critical
value for dynamo bifurcation, while maintaining a dipolar magnetic field.

\begin{figure*}
\centering
\includegraphics[width=7cm]{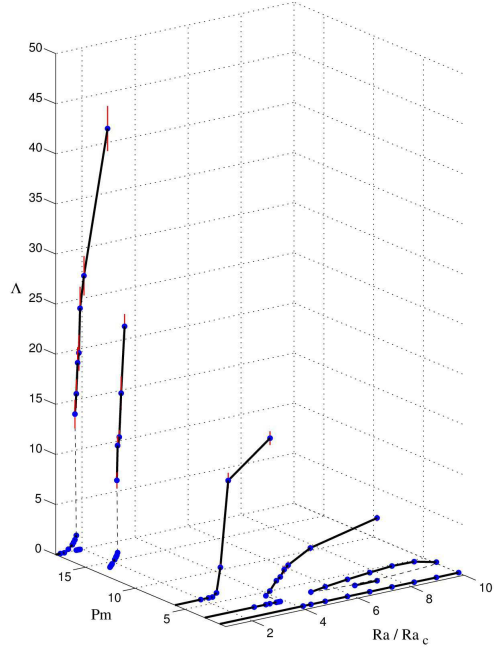}
\caption{Three-dimensional bifurcation diagram for a fixed Ekman number
$\Ek=3\cdot 10^{-4}$. Solid lines mark linear interpolation between
realised direct numerical simulations. Dashed lines offer a plausible interpretation of the unstable
branches.}
\label{fig3D}
\end{figure*} 

This new branch completes the sequence of bifurcation diagrams introduced
in \cite{MD09}, and the complete three dimensional bifurcation diagram 
\citep[including the results of][]{MD09} for 
$\Ek=3\, 10^{-4}$ and $\Pr =1$ is reported on Figure~\ref{fig3D} versus
$\Ray/\Ray_c$ and $\Pm$. {The corresponding two-dimensional bifurcation diagrams, for lower values of \Pm,
 are available in \cite{MD09}.}
{On such a three-dimensional diagram, the dynamo bifurcation can also be envisaged at fixed value of
$\Ray/\Ray_c$ 
and varying $\Pm$.}

Figure~\ref{fig3D} demonstrates
how the transition between the different types of bifurcation takes place
for different values of Pm. 
The study of \cite{MD09} indicates
that as E
is decreased (in the moderate range numerically achievable), the overall bifurcation diagram remains largely unaltered but
shifted towards lower values of Pm and larger $\Ray/\Ray_c$.

\begin{figure*}
\centerline{\includegraphics[width=\textwidth]{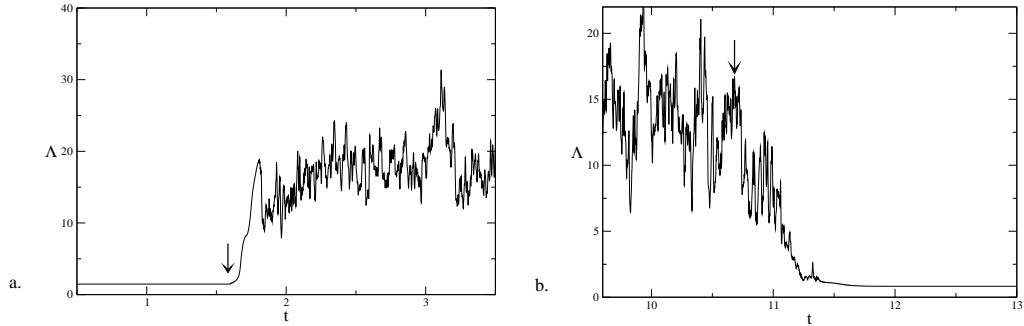}}
\caption{Transition between 
the weak- and strong-field branches for $\Ek=3 \cdot 10^{-4}$ and $\Pm=18$ (same as figure
\ref{fig_bif3a}.b). 
Runaway growth from the weak- to the strong-field branch as $\Ray/\Ray_c$
is increased (arrow) from
$1.78$ to $1.83$ (a) and catastrophic decay to the weak-field
branch as it is decreased (arrow) from $1.73$ to $1.68$
(b). The lower branch (left part of (a) and right part of (b)
corresponds to small, but non-vanishing magnetic fields.}
\label{fig_bif3b}
\end{figure*}

Transitions between these two branches of dynamo solutions are obtained by varying
only slightly the control parameter at the edge of a given branch.
This can produce either a runaway field growth (Fig~\ref{fig_bif3b}a), or a catastrophic
collapse (Fig~\ref{fig_bif3b}b) of the magnetic field. The time at which
the forcing (as measured by the Rayleigh number) has been modified (by less
than $3\%$ in each case) is indicated by an arrow on each graph.

{
No significant changes on the typical length scale of the flow can be
reported by comparing the weak- and strong-field branches.
This is probably due to the fact that the viscous length scale is not very
small at the value of the Ekman number considered here ($3\cdot
10^{-4}$).  Smaller values of the Ekman number are undoubtedly needed
if one is to appreciate a change in the typical length scale of the
flow.  One can note however that the Nusselt number is $80\%$ larger
on the strong-field branch than on the weak-field branch. 
}

\section{Force balance}

The bifurcation diagram presented on Figure~\ref{fig_bif3a} 
is reminiscent of a longstanding theoretical expectation originally
introduced as the `weak'- and `strong'-field branch~\citep[see][]{Roberts,RobertsS92}. 
The existence of these two branches in the limit of vanishing viscous
forces was introduced through the investigation of magneto-convection
studies \citep[see][for reviews]{Proctor94,Fearn86}.

\cite{Soward79} investigated the onset of magneto-convection in the cylindrical
annulus configuration with sloping boundaries. 
He found that in most cases the critical Rayleigh number
first starts to increase with the Elsasser number, until $\Elsclass \sim
\calO(\Ek^{1/3})$, before decreasing.
This pointed to the probable existence of a weak-field branch, and
the occurrence of a turning point marking the end
of the weak-field branch when $\Elsclass \sim
\calO(\Ek^{1/3})$.
Simultaneously, \cite{Fearn79a,Fearn79b}
performed a similar study in the spherical geometry. There again, the Rayleigh number for the thermal
Rossby mode may first increase with increasing Elsasser number, yet it
eventually decreases to reach a minimum for $\Elsclass \sim
\calO(1)$. A more recent study of magneto-convection \citep{Jonesetal2003}
focused on the ``weak-field'' regime and confirmed its existence.

The above asymptotic scenario assumes a small value of both E and Pm, whereas direct
numerical simulations in the self-excited dynamo regime require
overestimated values of both numbers.

\begin{figure}
\centerline{\includegraphics[width=8cm]{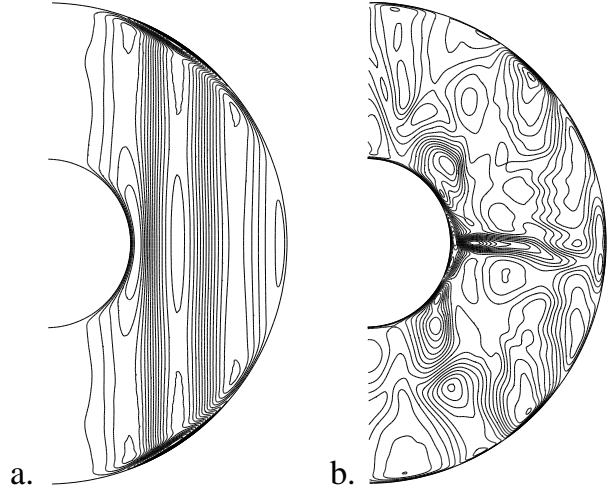}}
\caption{Azimuthal velocity in a meridional cross-section for
  an arbitrary time on the weak-field branch ({\bf a.}) and on the
  strong-field branch ({\bf b.}) for the same parameter set ($\Pm=18$, $\Ray/\Ray_c=1.73$).} 
\label{cross}
\end{figure}

In order to test the above ideas in the numerical simulations,
we need to investigate the dominant force balance
relevant to these dynamo modes. Figure~\ref{cross} presents an
instantaneous cross-section of the zonal velocity on both branches for a
given parameter set ($\Pm=18$, $\Ray/\Ray_c=1.73$).
The contour intervals are equally spaced between the minimum
  and the maximum value for each figure. The zonal flow is nearly
  three times larger on the left panel, so that the contour intervals
  are not identical on the two plots. 
On the one hand, the weak-field branch saturates while the zonal flow remains
essentially geostrophic; the flow is characterised by quasi-geostrophic
convection columns. 
The zonal flow on the strong-field branch, on the other hand, strongly
departs from bidimensionality, demonstrating that the rapid rotation
constraint has been relaxed.
On figure~\ref{cross}.b, a localised jet appears near the equator which marks a clear
departure from geostrophy.
{The flow is in general less anisotropic along the
direction of the axis of rotation.}
If this corresponds to the weak-field vs strong-field branches as
introduced by P.H.~Roberts, it implies that the Lorentz forced has achieved
a balance with the Coriolis term and thus relaxed the rapid rotation
constraint. {This dominant balance is usually referred to as the ``magnetostrophic balance''.}

The Elsasser number $\Elsclass$ was introduced to measure an
order-of-magnitude of the relative strength of the Lorentz force with
respect to the 
Coriolis force. It achieves this aim remarkably well in asymptotic studies,
for the huge distinction between the strong-field balance,
characterised by $\Elsclass \sim \calO (1) $ and the weak-field branch
characterised by $\Elsclass \sim \calO (\Ek ^{1/3}) $. 
In numerical works, however, as small parameters (such as the Ekman number) are not
asymptotically small, the measure provided by this non-dimensional number
is then not accurate enough. Finer estimates of this force balance can
then be constructed. Introducing $\left\{ \cdot \right\}$ as an ``order of
magnitude'' operator, we can write
\begin{equation}
\frac{\left\{ (\mu \rho)^{-1} \left(\bfnabla \times \bfB\right) \times \bfB \right\}}
{\left\{ 2 \, \Omega \times \bfu \right\}}
=
\frac{ B ^2}
{ 2 \, \Omega \, \mu \, \rho \, U \, \ell_B} \, ,
\label{Els_bal}
\end{equation} 
where $U$ is a typical, say root mean square (r.m.s.) value for the velocity field, $B$ a
typical value for the magnetic field, and $\ell_B$ the typical magnetic
dissipation length scale \citep[see also][]{OrubaD14a}.

The classical definition of the Elsasser number is obtained by assuming
$\ell_B \sim L$ and a statistical balance between induction and diffusion 
of the magnetic field
\begin{equation}
\left\{\bfnabla \times \left( \bfu \times \bfB \right) \right\}
\sim
\left\{\eta \bfnabla \times \bfnabla \times \bfB \right\} \, ,
\label{Elsass1}
\end{equation} 
which yields  
\(U \sim \eta / L\).
Then (\ref{Els_bal}) provides the standard expression
for the Elsasser number
\(
\Elsclass = {B ^2}/(2 \, \Omega \, \rho \, \mu \, \eta) \, .
\)
This expression provides a sensible description of the
force balance for asymptotic studies, yet finer estimates appear to be
needed for numerical studies.

\begin{figure}
\centerline{\includegraphics[width=9cm]{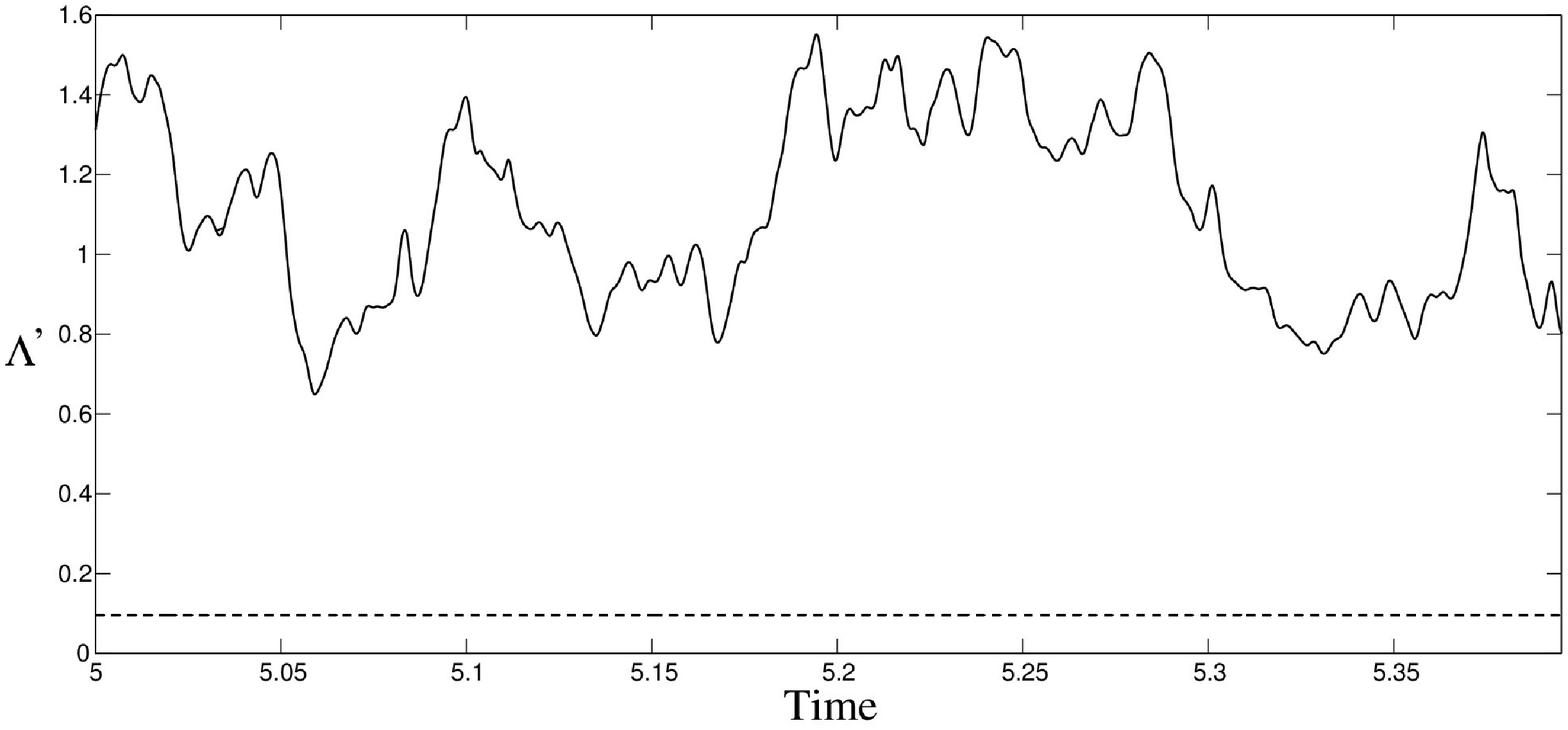}}
\caption{{Fluctuation of the modified Elsasser number on both
branches for the same parameter set ($\Pm=18$, $\Ray/\Ray_c=1.73$).}}
\label{Time_evol}
\end{figure}

One can note, for example that \(U \sim \eta / L\) amounts to assuming
Rm$\sim\calO(1)$. 
A finer description of the force balance {(\ref{Els_bal})} can be obtained
by estimating $U$  via \(\Rm \, \eta / L\).
Inserting this definition in (\ref{Els_bal}) yields
\begin{equation}
\Elsnum = 
\frac{B ^2 \, L}{2 \, \Omega \, \rho \, \mu \, \eta\, \Rm\, \ell_B}
= \Elsclass\, \frac{L}{\Rm\, \ell_B} \, .
\end{equation} 
Table~\ref{tab:Els} presents a comparison of the classical Elsasser number $\Elsclass$
and the modified Elsasser number $\Elsnum$ on both branches.
{The magnetic Reynolds number Rm is here defined on the RMS velocity, and the
typical magnetic
dissipation length scale $\ell_B$  is defined, as in \cite{OrubaD14a}, as
\begin{equation}
\ell_B ^2 = \frac{\int_V \bfB ^2 \dd V}{\int_V (\bfnabla \times \bfB ) ^2 \dd V} \, .
\end{equation}
}
{Figure~\ref{Time_evol} presents the time variation of the modified Elsasser 
number on the weak-field branch (dashed) and on the strong-field branch (solid line) 
for the same parameter set, $\Pm=18$, $\Ray/\Ray_c=1.73$.}

The modified
Elsasser number, offering a finer description of the force balance, reveals
that the Lorentz force is significantly weaker than the Coriolis force on
the weak-field branch and that the two terms are indeed of comparable amplitude on
the strong-field branch.

\begin{table}
  \begin{center}
\def~{\hphantom{0}}
  \begin{tabular}{lcccccccc}
Branch & $\Ray$ & $\Ray/\Ray_c$ & $\Pm$ & Rm & $\ell_B$ & $\Elsclass$ & $\Elsnum$\\
\\[3pt]
Weak   & 105 & 1.73 & 18 &  195 & 0.097 &  1.14  & 0.06\\
Strong & 105 & 1.73 & 18 &  145 & 0.082 &  13.6  & 1.14\\
Strong & 125 & 2.05 & 18 &  207 & 0.075 &  27.2  & 1.75\\ 
Strong & 112 & 1.84 & 14 &  150 & 0.083 &  11.05 & 0.89\\
  \end{tabular}
  \caption{Typical estimates of the Elsasser number and the modified
    Elsasser number ($\Elsnum = \Elsclass\, L/(\Rm\, \ell_B)$, see text) on
  the weak- and strong-field branches. The modified Elsasser number offers a
  finer measurement of the force balance.}
  \label{tab:Els}
  \end{center}
\end{table}

\begin{figure}
\centerline{\includegraphics[width=8.0cm]{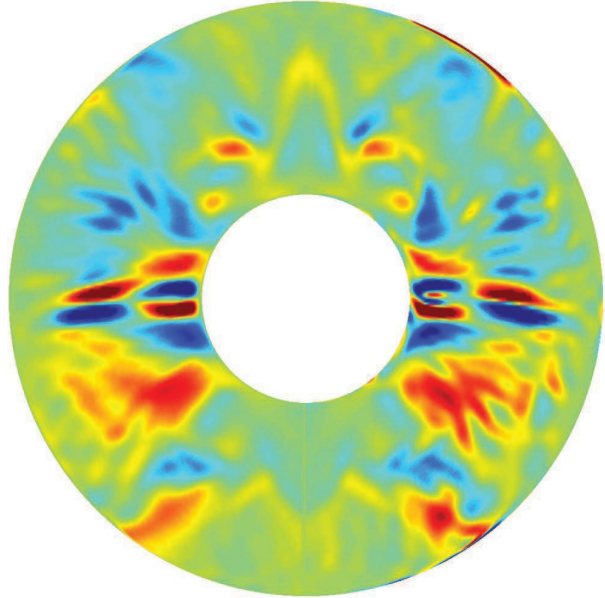}}
\caption{Meridional cross-sections on the strong-field branch 
at a given time for the same parameters and 
same phase as in figure~\ref{cross}.b.
On the left side 
$-2\, {\partial \bfu}/{\partial z}\cdot \bfe_r$ 
is presented, and on the right side $\left( \bfnabla \times\left( \left(\bfnabla \times
\bfB \right) \times \bfB\right)\right)\cdot \bfe_r$ using the same color range.}
\label{Taylor}
\end{figure}

The orders of magnitude derived above indicate that the anticipated balance
between the Coriolis and Lorentz forces is plausible. To achieve a finer
validation, than simple orders of magnitude, we can assess whether the two
terms tend to balance each other locally in space. To this aim, one can
consider the curl of the momentum equation (\ref{NS}), neglecting both the
inertial term 
and the viscous term, 
\begin{equation}
{-} 2\, \frac{\partial \bfu}{\partial z}
\sim \Ray \Rob \, \bfnabla \times (T \bfr) + \bfnabla \times\left( \left(\bfnabla \times
\bfB \right) \times \bfB\right)\, .
\end{equation} 
If we now consider the radial component of the above equation, i.e. its toroidal component, the first
term on the right-hand-side disappears. 
The remaining two terms were computed numerically at a given instant in time and
on a cross-section in an arbitrary meridional plane. These quantities are
presented on Figure~\ref{Taylor}.

Deviations between the two cross-sections can imply only non-vanishing inertial
and/or viscous effects. {Estimations of these terms reveals
  that the viscous term accounts for the differences visible on the
  figures (inertia being one order of magnitude smaller).}
 The comparison reveals such effects (in
particular in viscous boundary layers), but otherwise clearly demonstrates that the
radial component of the curl of the Lorentz force balances that of the
Coriolis force, as expected in the strong-field limit.

\begin{figure*}
\centerline{{\small a.}\includegraphics[width=0.43\textwidth]{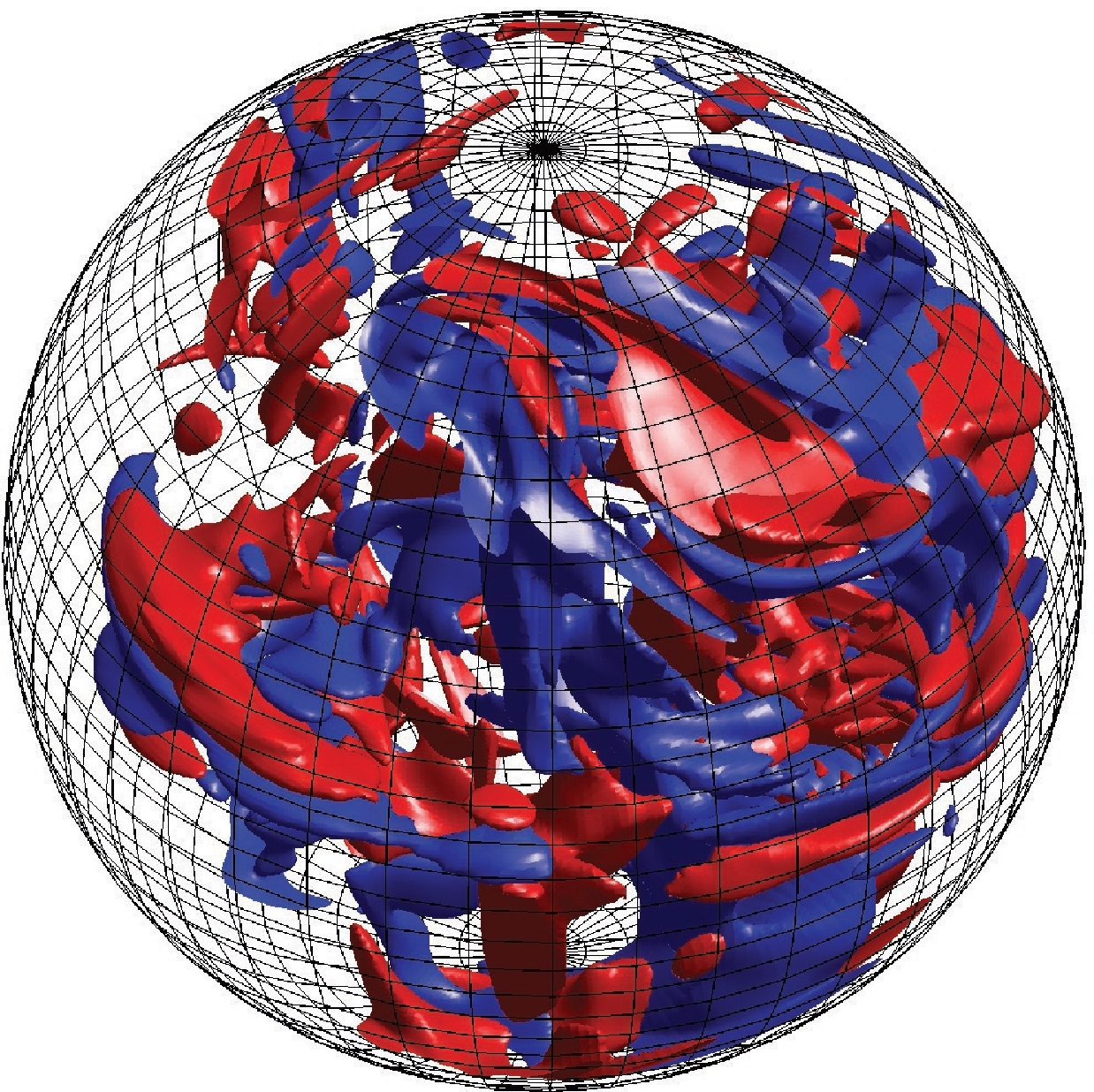}
\hskip6mm{\small b.}\includegraphics[width=0.43\textwidth]{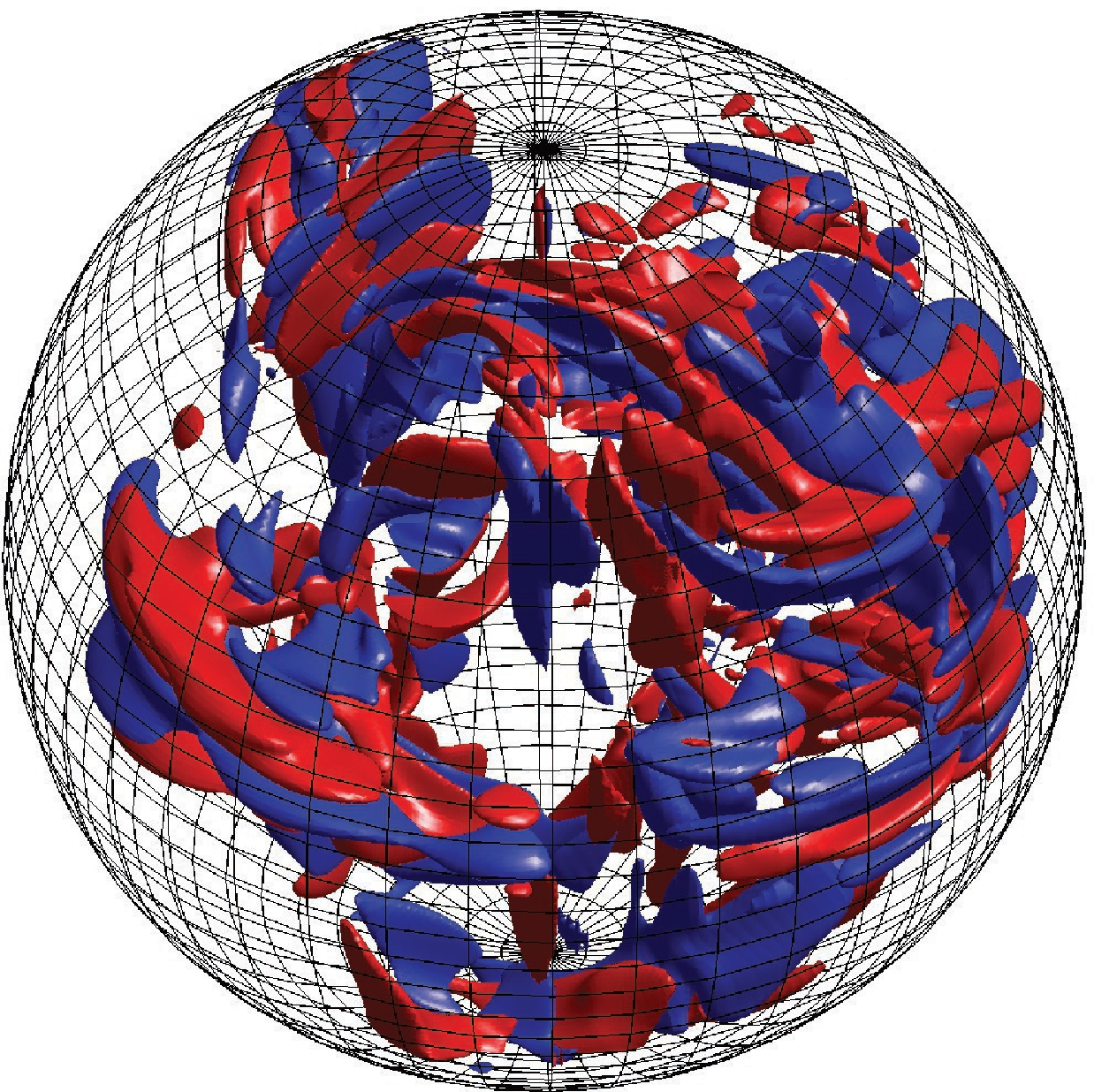}}
\caption{{Three dimensional iso-surfaces of the radial component of the curl of the Coriolis force (a.) and the Lorentz force (b.). }
}
\label{fig_Iso}
\end{figure*}

{The Viscous force will of course not always be negligible in
  the parameter regime considered here. It can be more important at
  some places or time. To illustrate this, the quantities represented
  on Figure~\ref{Taylor} are represented at a later time in the form
  of three-dimensional isosurfaces on 
  Figure~\ref{fig_Iso}. The blue and red isosurfaces
  respectively, correspond to $\pm90\%$  of the peak values.
  While deviations from magnetostrophy are obvious in particular
  comparing the centre of each figure, the dominant magnetostrophic balance is
  highlighted. Deviations are here primarily due to viscous
  forces. Boundary layers have not been represented in these figures.}

Figures~\ref{Taylor} and \ref{fig_Iso} highlight the balance between
  the non-gradient part of the Lorentz and Coriolis terms. Differences
  are primarily due to viscous effects and remain small (but non-vanishing) 
  on average. The r.m.s value of the sum of the two quantities
  plotted in these figures, averaged in time, exceeds by a factor 5
  that of their difference.

\section{Discussion}\label{sec:disc}

Numerical models of self-excited dynamos usually correspond to two distinct
branches, either viscous-dipolar or inertial-multipolar
\citep[e.g.][]{Kutzner,Jones11,OrubaD14b}.
This work introduces a third dynamo branch in the parameter regime which
is numerically achievable with present computational resources. 
{What if this new branch was relevant to the geodynamo?}

\subsection{Physical interpretation}
In numerical models, at moderate Ekman numbers, inertial
forces increase too rapidly with the control parameter (the Rayleigh number)
to allow for a strong-field branch balance (the magnetic field amplitude does not
increase rapidly enough, and inertial forces enter the dominant balance
before the Lorentz force). In order to observe at a given Ekman number the third branch
introduced in this work, one needs to increase the magnetic Prandtl
number in order to decrease the prefactor of inertial forces.
This results in a larger magnetic Reynolds number for a given value of the
Rayleigh number. 

Using large values of the magnetic Prandtl number thus allows for the
runaway field solution anticipated theoretically. {We observe that the} Lorentz force 
becomes large enough (while inertia remains small enough not to modify the
nature of dynamo action) in order for the Lorentz force to relax 
the constraints of rapid rotation.

\begin{figure*}
\centering
\includegraphics[width=9cm]{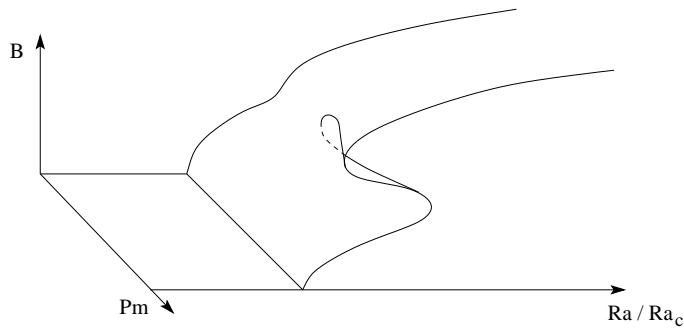}
\caption{{Sketch of a tentative two dimensional bifurcation diagram near the region 
of existence of the weak- and strong-field branches. The fold in the surface accounts for the observed two branches of solutions.}}
\label{figfolding}
\end{figure*} 

The three-dimensional bifurcation diagram (Figure~\ref{fig3D}) can be tentatively sketched near the region
in which the strong- and weak-field branches coexist (see Figure~\ref{figfolding}).
The change of branch as Ra is varied at fixed Pm corresponds to a fold of the surface of solutions.
It is clear from such a representation that, in the two dimensional parameter space (E, Pm), one can 
continuously move from the lower to the upper branch, avoiding the fold singularity. 
The corresponding intermediate models would presumably involve a continuous decrease of viscous forces.
These intermediate solutions (at lower values of Pm) will then have 
characteristics continuously varying from that of the weak-branch to that of the strong-branch. 
It follows that some numerical models obtained at lower values of Pm will necessarily have some 
characteristics of the strong-field branch. Such is the case in particular for models with a large 
magnetic Prandtl number 
(though not large enough for the bistability to occur) and large Rayleigh number (so that viscous 
effects are reduced), but not too large (to avoid the inertial, non-dipolar, branch).
In such a model, time-averaged force balance can tend to magnetostrophy 
\citep[e.g.][]{Aubert,Sreeni_rev}.

It should be stressed as well that large values of Pm have been studied in a few earlier numerical works
\citep[e.g.][]{Olson,Gubbins}, though without pointing to the existence of a weak- and a strong-field branch.

\subsection{Distinguished limit}

\begin{figure}
\centerline{\includegraphics[width=7.0cm]{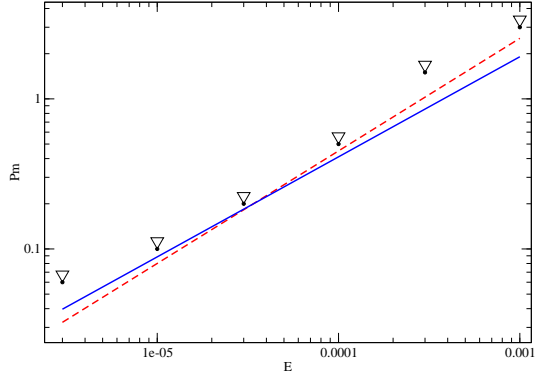}}
\caption{Minimal value of the magnetic Prandtl number $\Pm _c$ below which
  the dipolar viscous branch is lost for a given Ekman number. Triangles
  correspond to numerical data produced by \cite{ChristensenA06}, and the
  dashed line to their empirical fit ${\rm Pm}_c \sim {\rm E}^{3/4} $.
The solid line corresponds to the ${\rm Pm_c} \sim {\rm E}^{2/3} $
scaling 
\citep{DJLLM}.}
\label{CA06}
\end{figure}
We have seen in \cite{MD09} that the nature of the dynamo bifurcation
strongly depends on the parameters (in particular $\Pm$ at fixed
$\Ek$).
The current approach to geodynamo modelling consists either in trying to 
explore the whole range of magnetic Prandtl number at fixed
Ekman number, or, too often, in trying to decrease the value of the
magnetic Prandtl number as low as possible for a given Ekman number 
(the viscous dipolar solution being lost for $\Pm$ lower than a critical
value $\Pm_c$).
The present work {suggests} that, in order to preserve the relevant
force balance, both $\Ek$ and $\Pm$
being small parameters in the Earth's core, 
they {could} be related in numerical studies via a
distinguished limit, involving only one small parameter $\varepsilon$.
Ideally the nature of the
dynamo bifurcation should be preserved in the limiting process. 

In order to propose such a scaling, one could be guided by the scaling for the minimum magnetic
Prandtl number $\Pm_c$ as a function of the Ekman number
$\Ek$. {This relations stems from numerical simulations in the viscous branch: it
is however the only regime that has been widely covered in numerical simulations.} 
The available data are illustrated in Figure~\ref{CA06}. Decreasing $\Pm$ at otherwise
fixed parameters amounts to decreasing the magnetic Reynolds number.
The
dipolar viscous branch is thus lost for $\Pm < Pm_c$
which decreases with decreasing values of the Ekman number
\citep{Kutzner,ChristensenA06}.
It follows a scaling of the form
\(
\Pm_{c} \sim \Ek^{\alpha} \, ,
\)
in which $\alpha$ needs to be determined. 
\cite{ChristensenA06} proposed an empirical fit 
with $\alpha=3/4$. \cite{DJLLM}
proposed on the basis of exponential growth associated with a locally
time dependent shear, a scaling of the form $\alpha=2/3$, which seems to
match the numerical data equally well and is 
guided by a plausible argument.
For simplicity, we will use this latter scaling for illustration
purposes below, but the same reasoning would apply with a different
exponent.

In order to introduce a single small parameter $\varepsilon$ to control both
quantities, one can write
\(
\Ek \sim A \, \varepsilon^B
\)
and
\(
\Pm \sim C \, \varepsilon ^D \, .
\)
Without loss of generality, one can set $A=1$, up to a redefinition of
$\varepsilon$. In order to preserve the nature of the dynamo bifurcation in
the limiting process, we propose 
\(
\Pm ^3  \sim \Ek^2 \, ,
\)
and it follows that 
\begin{equation}
\Ek \sim \varepsilon^3  \qquad \Pm\sim C \, \varepsilon ^2 \, .
\label{constantC}
\end{equation} 
This distinguished limit ensures that both $\Ek$ and $\Pm$ tend to zero
with $\varepsilon$ and that they are related in such a way that the nature
of the solution, i.e. its dynamo property, should be preserved in the
limiting process. The coefficient $C$ can be estimated via
sensible estimates for the Earth's core, such as $\Ek \simeq 10^{-14}$ and $\Pm \simeq 10^{-6}$: the first
equation naturally provides $\varepsilon \simeq 2\cdot 10^{-5}$, while the second
yields $C \simeq 2\cdot 10^3$, which is a rather large prefactor. 
Applying (\ref{constantC}) to the numerical models presented in this work
(with $\Ek=3 \cdot 10^{-4}$), in turn yields 
$\Pm \simeq 10$. 
{This distinguished limit thus yields values of the magnetic
  Prandtl number larger than unity (similar to those used in our
  numerical studies) for the Ekman number investigated here.}

The idea behind such distinguished limits is not to aim at a
large magnetic Prandtl number limit, as both the Ekman number and the Prandtl number vanish
asymptotically in the limiting process. For the moderate values of the
small parameter $\varepsilon$,
achievable with current computational resources, the proposed distinguished
limit however suggests that the use of values of $\Pm$ larger than unity is relevant. 

As computational resources increase, one should be able in the near
future to investigate the behaviour of this strong-field branch for
lower values of the Ekman number, and thus lower values of the
magnetic Prandtl number. This will allow investigation of interesting and important
issues in particular on the evolution of the relevant length scale as
the magnetic Prandtl number becomes less than unity.

\section{Conclusions}\label{sec:concl}
A dominant magnetostrophic balance can be
established in direct numerical simulations of rotating spherical
dynamos.
Magnetostrophy is not satisfied everywhere and for for all time.

The weak- and strong-field branches anticipated from asymptotic
studies of magneto-convection {are approached} in direct numerical simulations 
for some parameter values.
In order for inertial forces to be small enough to allow this regime, it is
necessary to relate the magnetic Prandtl number to the Ekman number in the
form of a distinguished limit.
Further studies will need to decrease the Ekman number to ensure a
clear distinction between the small scale flow on the viscous branch
and the large-scale flow on the strong-field branch.
The next important challenge for direct numerical models would be to
maintain dynamo action for $\Ray/\Ray_c <1$ as expected theoretically.
The role of the Prandtl number (fixed to unity here) in controlling the relative strength of 
the advection versus diffusion of heat also deserves further studies.
{Further numerical studies of this branch could include varying Pr. For
instance the limit of large Pm with small q would allow for significant
nonlinearities in the energy equation, while controlling the amplitude of inertial effects.}

\begin{acknowledgments}
The author is very grateful to Vincent Morin, Ludovic Petitdemange and
Ludivine Oruba for their help at various stages of the preparation of this article.
Stephan Fauve provided useful comments on earlier versions of this work.
Computations were performed on the CEMAG parallel computer
{and on the HPC resources of MesoPSL, financed
by the Region Ile de France and the project Equip@Meso (reference
ANR-10-EQPX-29-01) of the programme Investissements d'Avenir supervised
by the Agence Nationale pour la Recherche.}
\end{acknowledgments}

\end{document}